\documentclass{PoS}

\usepackage{cite}
\usepackage{amssymb}
\usepackage{amsmath}

\newcommand{\captionit}[1]{\caption{\textit{#1}}}
\def\proc{$pp\to e^+\nu_e \mu^-\bar{\nu}_\mu b\bar{b}j + X$}
\def\gev{\text{~GeV}}
\def\tev{\text{~TeV}}
\def\helacnlo{\textsc{Helac-Nlo}}
\def\helacdip{\textsc{Helac-Dipoles}}
\def\helacloop{\textsc{Helac-1Loop}}
\def\phegas{\textsc{Helac-Phegas}}
\def\kaleu{\textsc{Kaleu}}
\def\cuttools{\textsc{CutTools}}
\def\oneloop{\textsc{OneLOop}}
\def\toppp{\textsc{Top++}}

\title{NLO QCD off-shell effects for top pair production with a jet in the dilepton channel}

\ShortTitle{NLO QCD off-shell effects for $t\bar{t}j$ production in the dilepton channel}

\author{\speaker{Manfred Kraus}\thanks{Preprint number: TTK-16-29}\\
        Institute for Theoretical Particle Physics and Cosmology\\
        RWTH Aachen University, D-52056 Aachen, Germany \\
        E-mail: \email{kraus@physik.rwth-aachen.de}}

\abstract{A short summary of a NLO QCD calculation for the top-quark pair production with a jet in the dilepton decay channel is presented.
          The calculation takes into account all double, single, non-resonant and interference contributions. Thus, NLO QCD corrections
          are computed for the process \proc{}. This calculation provides the first complete description, which overcomes the narrow width approximation
          for $t\bar{t}j$ process in the dilepton decay channel.
          Results for the total cross section and differential distributions for the LHC at $\sqrt{s}=8\tev$
          are shown.}

\FullConference{Fourth Annual Large Hadron Collider Physics\\
		13-18 June 2016\\
		Lund, Sweden}

\begin{document}

\section{Introduction}
After the discovery of the top quark at the Tevatron in 1995, its properties have been extensively measured ever since.
The LHC then overtook the top-quark physics program after its commissoning in 2009.
In most cases the LHC Run 1 results obtained at $\sqrt{s}=7,8\tev$ already improve the Tevatron results. However, with the
start of Run 2 at the center-of-mass energy $\sqrt{s}=13\tev$, the precision era for top-quark physics has begun. Important measurements are
the determination of the top-quark coupling to gauge bosons and the Higgs boson, spin-correlations of the top-quark decay products as well as
fiducial cross sections and differential distributions.
For now, the most advanced calculations for stable top-quark pair production are the NNLO+NNLL prediction~\cite{Czakon:2013goa} of the total
cross section and NNLO accurate prediction for differential distributions~\cite{Czakon:2015owf,Czakon:2016dgf}. However, top-quarks are extremely
unstable particles, and therefore decay before they can hadronize. Thus, top-quarks can be studied directly from its decay products, where in the
Standard Model they nearly in all cases decay into a bottom quark and a $W$ boson. Depending on the further decay of the $W$ boson one can 
classify the decay modes into leptonic, $W\to \ell\nu_\ell$, or hadronic, $W\to q\bar{q}^\prime$, decays. The cleanest signature is obtained if 
both top quarks decay leptonically and therefore the final-state consist out of two bottom-quark induced jets, two oppositely charged leptons and 
missing energy. Top-quark decays were included in the NLO calculations within the narrow width approximation in the works of Refs.~\cite
{Melnikov:2009dn,Campbell:2012uf}. Another approach was taken in  Refs~\cite{Denner:2010jp,Bevilacqua:2010qb,Denner:2012yc,Cascioli:2013wga,Frederix:2013gra,Heinrich:2013qaa} 
where top-quarks are not treated as on-shell particles anymore. Instead the leptonic decays have been included and NLO QCD corrections were 
calculated for the complete $pp\to e^+\nu_e \mu^- \bar{\nu}_\mu b\bar{b} +X$ process. Just recently also electroweak corrections were presented 
in Ref.~\cite{Denner:2016jyo} and a consistent matching to parton showers has been presented in Ref.~\cite{Jezo:2016ujg}.

\begin{figure}
\begin{center}
  \includegraphics[width=0.65\textwidth]{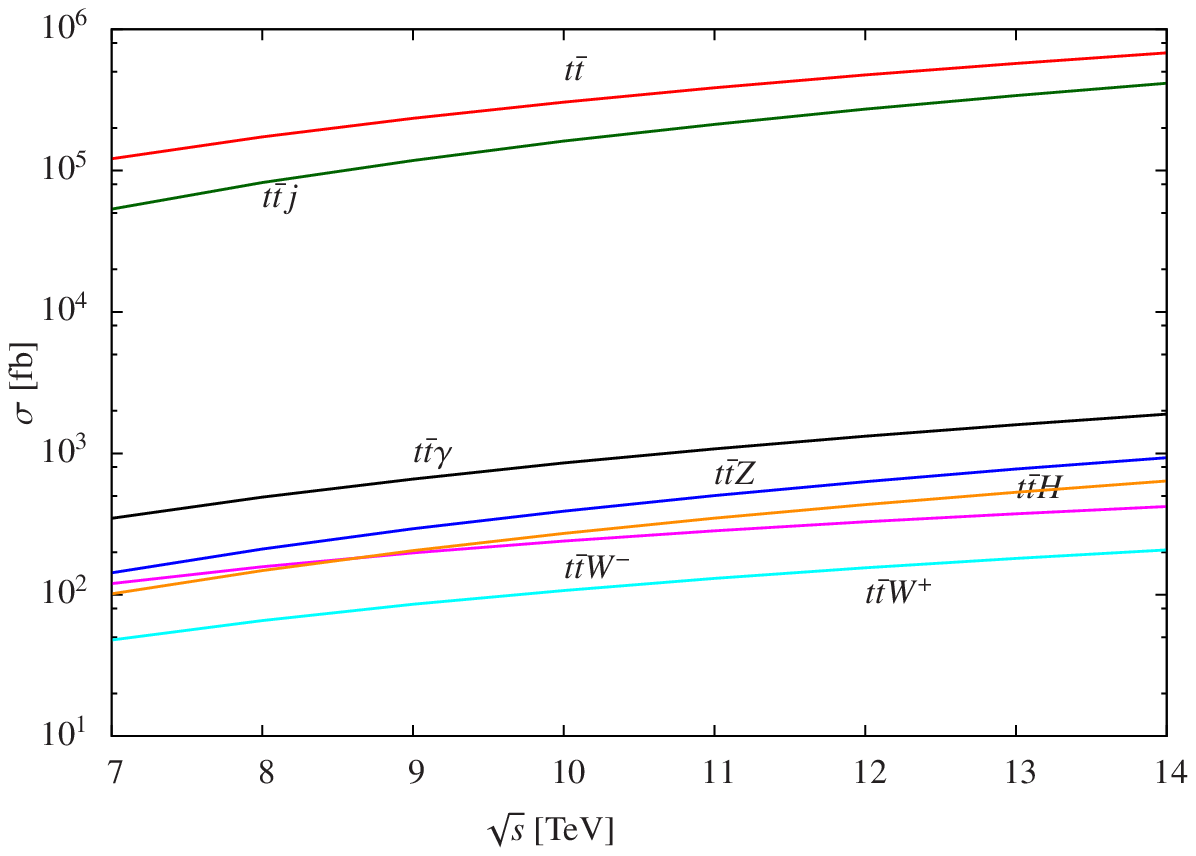}
  \captionit{The total cross sections for $t\bar{t}X$ processes as a function of $\sqrt{s}$.}
  \label{Fig:propaganda}
\end{center}
\end{figure}

At the very large energies of the Large Hadron Collider top-quark pairs are often observed accompanied by additional jets and gauge bosons.
In Fig.~\ref{Fig:propaganda} the total cross sections for the associated production processes of top-quark pairs are shown as a function of the 
center-of-mass energy $\sqrt{s}$.
It is clear, that the $t\bar{t}j$ production is the largest contribution to the inclusive $t\bar{t}$ sample for moderate values of the $p_T$ cut 
on the hardest jet.
 For a better estimate of this contribution, we show in Table~\ref{Tab:Xsecs} the NLO QCD cross sections for several different 
choices of the $p_T$ cut and the ratio to the NNLO $t\bar{t}$ cross section.
\begin{table}[h]
\begin{center}
\begin{tabular} {|c|r|c|}
  \hline
  $p_T(j_1)~[\gev]$ & $\sigma_{t\bar{t}j}^{NLO}~[pb]$ & $\sigma_{t\bar{t}j}^{NLO}/\sigma_{t\bar{t}}^{NNLO}~[\%]$ \\
  \hline 
  $40$ & $296.97 \pm 0.29$ & $37$ \\
  $60$ & $207.88 \pm 0.19$ & $26$ \\
  $80$ & $152.89 \pm 0.13$ & $19$ \\
  $100$ & $115.60 \pm 0.14$ & $14$ \\
  $120$ & $89.05 \pm 0.10$ & $11$ \\
  \hline
\end{tabular}
\captionit{The total cross section of $t\bar{t}j$ at NLO as a function of the jet $p_T$ cut.}
\label{Tab:Xsecs}
\end{center}
\end{table}
The values were obtained for $m_t=173.2\gev$ for the LHC at $\sqrt{s}=13\tev$ using the CT14 PDF sets~\cite{Dulat:2015mca}. NLO results were 
calculated using the \helacnlo{} framework~\cite{Bevilacqua:2011xh} while the NNLO prediction were obtained using \toppp{}~\cite{Czakon:2011xx}. 
For a cut of $p_T(j_1)>40\gev$ nearly $40\%$ of all $t\bar{t}$ pairs are accompanied by an additional hard jet. Consequently, precise theoretical
predictions for the $t\bar{t}j$ production are necessary. 

The process $pp\to t\bar{t}j + X$ has several important applications. First of all, the process can be utilitized in the determination of the 
top-quark mass parameter using an alternative method~\cite{Alioli:2013mxa} which has been already used by both collaborations~\cite{Aad:2015waa,CMS:2016khu}. 
On the other hand, $t\bar{t}j$ can be used to test physics beyond the Standard Model, for example, models with anomalous couplings to the top 
quark~\cite{Gresham:2011dg} would lead to resonances in the invariant mass spectrum of the top-quark and the additional jet. Finally, $t\bar{t}j$ 
constitutes the dominant background for Higgs boson production in the vector boson fusion process (VBF), $pp\to Hjj \to W^+W^-jj$, if the Higgs 
decays into a pair of $W$ bosons~\cite{Rainwater:1999sd,Kauer:2000hi}. 
\begin{figure}
\begin{center}
  \includegraphics[width=0.48\textwidth]{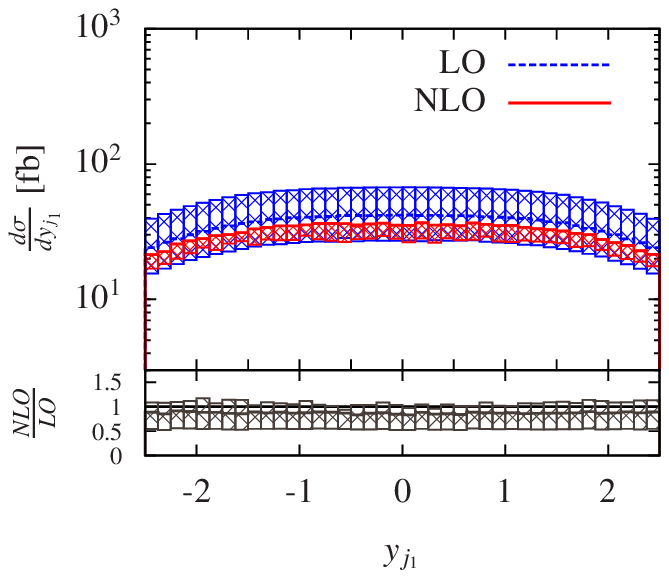}
  \includegraphics[width=0.48\textwidth]{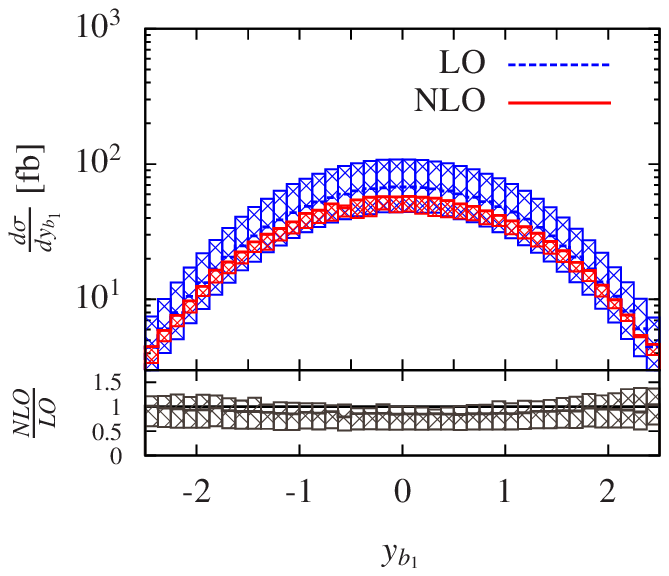} \\
  \captionit{Differential distributions for the rapidity of the hardest light jet (left) and b-jet (right).}
  \label{Fig:rapidity}
\end{center}
\end{figure}
For the VBF process the dominant kinematic signature are two forward-backward jets, which are required to be well separated in rapidity, i.e. $|y_1 - y_2| > 4$ 
and at the same time lie in opposite hemispheres of the detector, $y_1\times y_2 < 0$. In Fig.~\ref{Fig:rapidity} the rapidity of the hardest light and 
bottom-quark induced jet are depicted. In the case of top-quark pair production the two leading jets are the b-jets from the decay and as can be 
seen from Fig.~\ref{Fig:rapidity} these are produced very centrally in rapidity. Thus, top-quark pair production is strongly suppressed by the 
aforementioned selection cuts. On the other hand, if the top-quark pair is accompanied by an additional jet then the suppression due to phase space 
cuts are less severe than in the case of $t\bar{t}$ production. The reason is that the light jet can be one of the leading two jets and in 
addition it is produced very broad in rapidity as can be seen from Fig.~\ref{Fig:rapidity}. Thus this light jet circumvents the strong  
suppression and makes $t\bar{t}j$ production the dominant background.
\begin{figure}
\begin{center}
  \includegraphics[width=\textwidth]{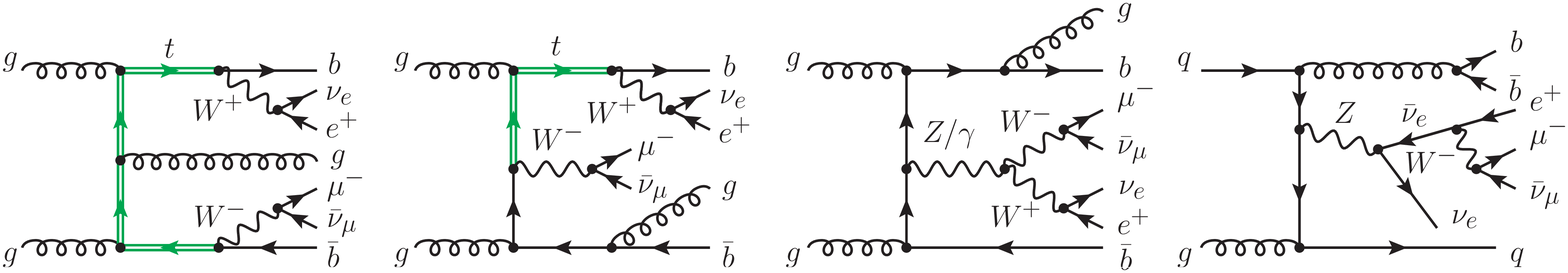}
  \captionit{Differential distributions for the rapidity of the hardest light jet (left) and b-jet (right).}
  \label{Fig:fd}
\end{center}
\end{figure}

In conclusion, precise theoretical predictions for $pp\to t\bar{t}j+X$ are mandatory. For stable top-quarks the NLO QCD corrections have been 
presented in Refs.~\cite{Dittmaier:2007wz,Dittmaier:2008uj} and were later supplemented by LO decays~\cite{Melnikov:2010iu} in the narrow width 
approximation. Later the top-quark decays have been extended to NLO accuracy~\cite{Melnikov:2011qx}. Top-quark pair production in association 
with one hard jet has also been matched to parton showers by several groups~\cite{Kardos:2011qa,Alioli:2011as,Czakon:2015cla}.
Recently, also the approximation of on-shell top-quark decays has been dropped and the calculation of the full process \proc{} has been presented 
in Ref.~\cite{Bevilacqua:2015qha}. Representative Feynman diagrams for the leading-order production at $\mathcal{O}(\alpha_s^3\alpha^4)$ are 
shown in Fig.~\ref{Fig:fd}. This calculation takes into account all double, single and non-resonant top-quark contributions as well as 
interferences between them. Additionally, top quarks and $W$, and $Z$ bosons are treated as unstable intermediate particles. Off-shell effects 
for the top quark are small for inclusive observables and of the order of $\Gamma_t/m_t \approx 1\%$, as has been found in various calculations~
\cite{Denner:2010jp,Bevilacqua:2010qb,Denner:2012yc,Frederix:2013gra,Cascioli:2013wga,Heinrich:2013qaa,Denner:2015yca,Bevilacqua:2015qha}. 
However, for certain observables off-shell effects can have a larger impact and can affect searches for the Higgs boson~\cite{AlcarazMaestre:2012vp} .
\section{Calculational framework}
The calculation of the next-to-leading order QCD corrections for the process\\ \proc{} is performed using the \helacnlo{} framework, which
has been already successfully utilized for a wide range of top-quark studies~\cite{Bevilacqua:2009zn,Bevilacqua:2010ve,Bevilacqua:2011aa,Worek:2011rd,
Bevilacqua:2012em,Bevilacqua:2014qfa}. The \helacnlo{} framework extends the \phegas{} Monte-Carlo generator~\cite{Kanaki:2000ey,Cafarella:2007pc} by
the capability of calculating NLO QCD corrections. It consists of two main packages:

\helacdip{}~\cite{Czakon:2009ss} for the evaluation of the real
corrections, where currently two subtraction schemes are available, namely the Catani-Seymour~\cite{Catani:1996vz,Catani:2002hc} and the
Nagy-Soper~\cite{Bevilacqua:2013iha} subtraction scheme. The complicated phase space integrations are performed using \kaleu{}~\cite{vanHameren:2010gg}.
The \helacloop{} package~\cite{vanHameren:2009dr} on the other side deals with the automated computation of the virtual corrections. The program
is based on the so-called OPP reduction method~\cite{Ossola:2006us,Ossola:2008xq,Draggiotis:2009yb} as implemented in \cuttools{}~\cite{Ossola:2007ax}
together with Dyson-Schwinger recursions for the computation of tree-level diagrams. Scalar one-loop integrals are computed using the \oneloop{}~\cite{vanHameren:2010cp}
library.

Treating the top quark as an intermediate particle gives rise to singularities due to on-shell propagators. To regularize resonant propagators the
complex-mass scheme~\cite{Denner:2005fg} is employed. Here, the mass of the top quark is replaced by $\mu_t^2 = m_t^2 -im_t\Gamma_t$, where $\Gamma_t$ is
the top-quark decay width.

\section{Numerical results}
\begin{figure}
\begin{center}
  \includegraphics[width=0.5\textwidth]{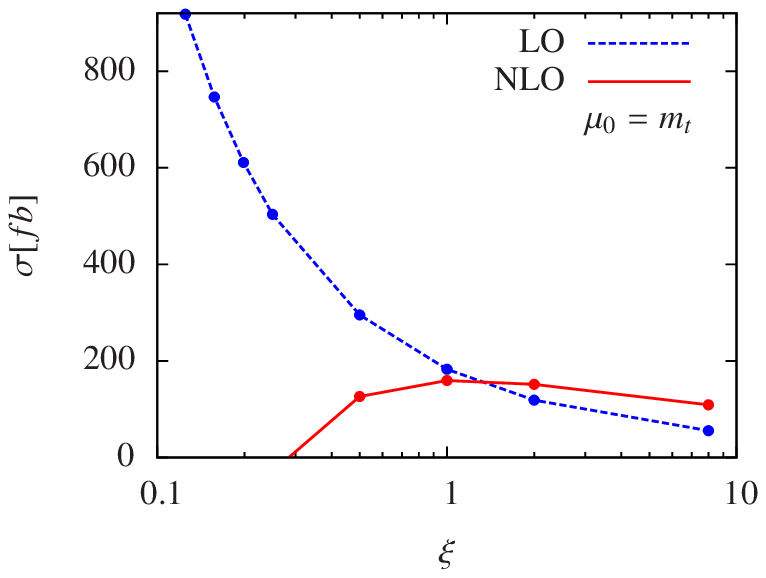}
  \captionit{The total cross section as a function of the common renormalization and factorization scale $\mu_R=\mu_F=m_t$.}
  \label{Fig:scaledep}
\end{center}
\end{figure}

\begin{figure}
\begin{center}
  \includegraphics[width=0.48\textwidth]{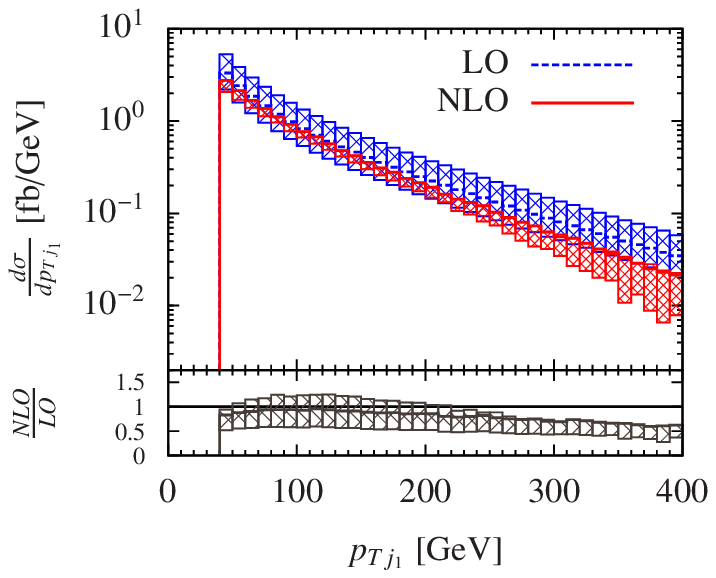}
  \includegraphics[width=0.48\textwidth]{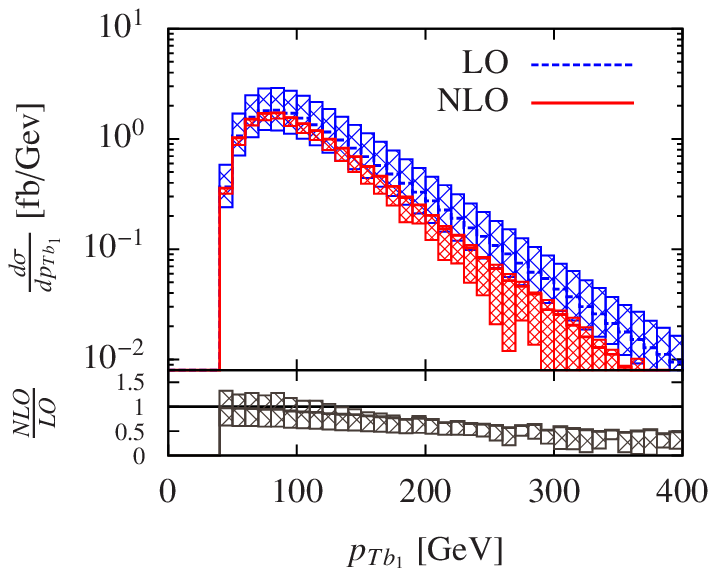} \\
  \includegraphics[width=0.48\textwidth]{hist_yj1.eps}
  \includegraphics[width=0.48\textwidth]{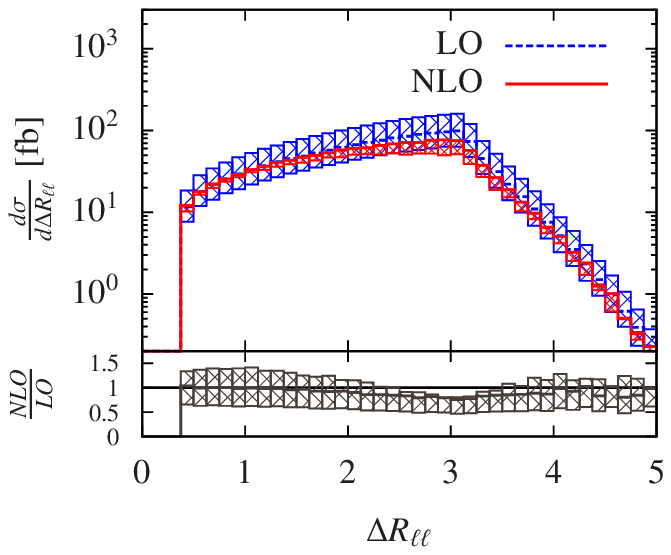} \\
  \includegraphics[width=0.48\textwidth]{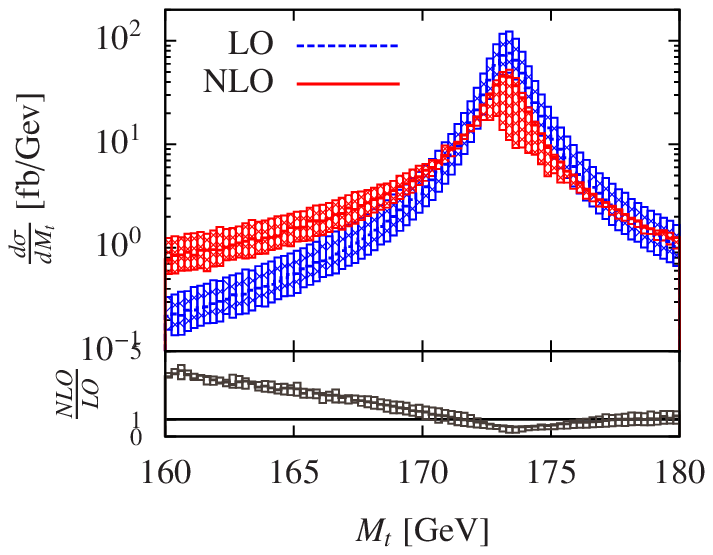}
  \includegraphics[width=0.48\textwidth]{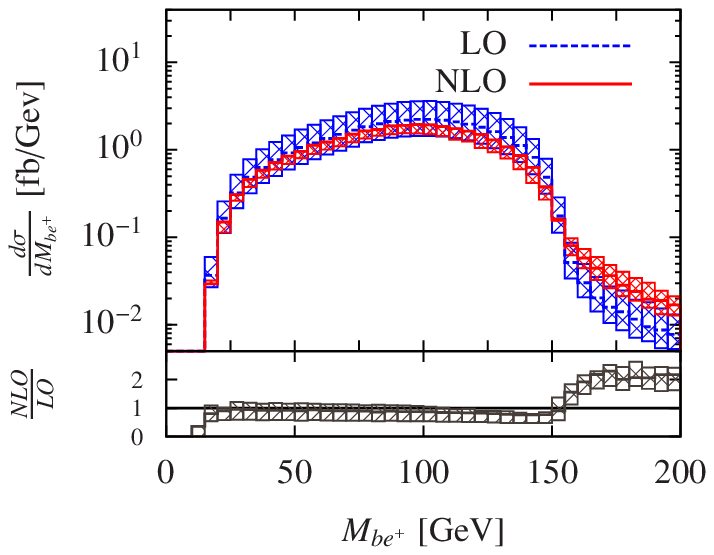}
  \captionit{Differential distributions as a function of the transverse momentum of the hardest jet $p_{T\;j_1}$, and of the hardest bottom-quark induced jet $p_{T\;,b_1}$, the rapidity
  of the hardest jet $y_{j_1}$, the angular separation $\Delta R_{\ell\ell}$ between the leptons, the averaged reconstructed invariant mass of the top quark and the invariant mass
  of the positron and the b-jet for \proc{} for the LHC at $\sqrt{s}=8\tev$. The uncertainty bands are obtained by scale variation by a factor of two. The lower panel shows the
  ratio of NLO to LO prediction.}
  \label{Fig:diffs}
\end{center}
\end{figure}

The following Standard Model input parameters are used to obtain the numerical results presented in this section
\begin{equation}
\begin{split}
  G_F &= 1.16637\cdot 10^{-5}\gev^{-2}\;, \qquad m_t = 173.3\gev\;,\\
  m_W &= 80.399\gev\;, \qquad \qquad\quad \Gamma_W = 2.09774\gev\;, \\
  m_Z &= 91.1876\gev\;, \qquad\qquad\quad \Gamma_Z = 2.50966\;, \\
  \Gamma_t^{LO} &= 1.48132\gev\;, \qquad\qquad \Gamma_t^{NLO} = 1.3542\gev\;.
\end{split}
\end{equation}
Predictions are given for the LHC working at a center-of-mass energy of $\sqrt{s}=8\tev$. The parton distribution functions togehter with the
corresponding one or two-loop running of $\alpha_s$ are taken from the \textsc{MSTW2008LO68cl} and \textsc{MSTW2008NLO68cl} PDF sets~\cite{Martin:2009iq} repsectively.
The factorization and renormalization scales are fixed to a common value of $\mu_R=\mu_F=m_t$. Bottom-quark induced scattering channels are neglected due to the strong
suppression of the PDFs. Jets are defined using the anti-$k_T$ jet-algorithm~\cite{Cacciari:2008gp} with a separation parameter of $R=0.5$, where only partons with pseudorapidity
$|\eta| < 5$ are considered by the jet-algorithm.
The final state is required to consist of two charged leptons, exactly two bottom-quark induced jets, at least one light jet and missing transverse momentum, where
additional phase space selection cuts are also imposed. For both, bottom-quark induced and light jets, commonly denoted with $j$ in the following, we require
\begin{equation}
  p_T(j) > 40\gev\;, \quad |y(j)| < 2.5\;, \quad \Delta R_{jj} > 0.5\;.
\end{equation}
While for the leptonic final-state particles we require
\begin{equation}
  p_T(\ell) > 30\gev\;, \quad |y(\ell)| < 2.5\;, \quad \Delta R_{\ell j} > 0.4\;, \quad \Delta R_{\ell\ell} > 0.4\;, \quad p_T^{miss}>40\gev\;.
\end{equation}
The cross section corresponding to the aforementioned setup is depicted in Fig.~\ref{Fig:scaledep} as a function of the renormalization and factorization scales.
The central prediction including the theoretical uncertainty, obtained by scale variation by
a factor of two, is given by
\begin{equation}
 \sigma^{LO} = 183.1^{+112.2~(+61\%)}_{-64.2~(-35\%)}~\text{fb}\;, \qquad \sigma^{NLO} = 159.7^{-33.1~(-21\%)}_{-7.9~(-5\%)}~\text{fb}\;.
\end{equation}
The NLO corrections induce a strong reduction of the residual scale dependence which can be seen in Fig.~\ref{Fig:scaledep}. For the total cross section the theoretical uncertainty
is reduced from $61\%$ at LO down to $21\%$ at NLO, by nearly a factor of three. At the same time, the higher-order corrections amount to a reduction of the total cross section
by $-13\%$.
Now turning to differential distributions; in Fig.~\ref{Fig:diffs} three pairs of differential distributions are depicted which pairwise share common features. In each plot the
leading-order (blue-dashed) and the next-to-leading order (red-solid) prediction is shown together with their corresponding theory uncertainties. The latter are obtained by varrying the
scale and choosing bin-by-bin the minimum and maximum value from the following set $\{m_t/2, m_t, 2m_t\}$. The lower panel depicts the differential K-factor.

The first pair of distributions to be discussed are the transverse momentum of the hardest jet $p_{T, j_1}$, and of the hardest bottom-quark induced jet $p_{T, b_1}$.
Both observables are dimensionful, where for example the beginning of the spectra are sensitive to the $t\bar{t}j$ threshold, while at the end of the plotted range the dominant
phase space regions are very hard jet configurations. For the transverse momentum of the hardest jet corrections are non uniform over the whole range and the LO shape can not be
rescaled by a constant factor. Furthermore, corrections to the shape of the order of $-50\%$ are found towards the end of the plotted range, while they are moderate at the beginning
of the spectrum.
For the transverse momentum of the hardest b-jet $p_{T, b_1}$ similar corrections have been found. Also here the LO shape is not rescaled by an overall factor and
shape distortions up to $-50\%$ in the tail are found. The common feature of both observables is that the
perturbative convergence is bad, which can be seen from the fact that in the tails the NLO predictions lie outside of the LO uncertainty band. Thus the LO uncertainty gives an
unreliable estimation of the missing higher-order corrections. To resolve this issue dynamic factorization and renormalization scales have to be chosen, which take into account
the transverse momentum of the leading jet.

The next two distributions are the rapidity of the hardest jet, $y_{j_1}$ and the angular separation of the leptons, $\Delta R_{\ell\ell}$.
These two observables represent angular distributions, which should converge faster than dimensionful observables, since the threshold region contributes to the whole range.
For the rapidity we find moderate and negative corrections to the overall normalization of the distribution. Additionally, the convergence in this particular distribution is
excellent.
The angular separation of the lepton pair is given by
\begin{equation}
  \Delta R_{\ell\ell} = \sqrt{\Delta\Phi_{\ell\ell}^2 + \Delta y_{\ell\ell}^2}\;,
\end{equation}
where $\Delta\Phi_{\ell\ell}$ is the difference in the azimuthal angles and $\Delta y_{\ell\ell}$ the difference in rapidities of the leptons.
The depicted distribution peaks approximately at $\Delta R_{\ell\ell} = 3$, which means that the leptons are produced mostly back-to-back.
The QCD corrections are quite flat over the whole region with the exception of the peak-region, where moderate negative corrections are found.

The remaining two observables to be discussed are the reconstructed invariant top-quark mass and the invariant mass of the positron and the b-jet.
Both observables highlight the included off-shell effects in this calculation. For instance, the reconstruction of the invariant top-quark mass
by using $M_t = \sqrt{p_t^2}$, where the top four-momentum is given by $p_t=p_b+p_\ell+p_\nu$, shows a clear Breit-Wigner shape. The NLO
corrections introduce large radiative corrections below the peak, which are generated by further radiation from the bottom quark which is not
recombined with its mother particle by the jet-algorithm.
The last observable, $M_{be^+}$, is phenomenologically important for top-quark mass parameter extraction due to its high sensitivity on the top-quark mass.
At the same time, the $M_{be^+}$ spectrum shows features from off-shell effects. The observable is defined by using the $b$-jet that minimizes the
invariant mass with the positron. If the same spectrum would be calculated at LO using the narrow width approximation then the phase space of the
top decay products is more constrained. For example, for on-shell top quarks and $W$ bosons, the invariant mass $M_{be^+}$ exhibits an upper boundary given by
\begin{equation}
  M_{be^+} = \sqrt{m_t^2-m_W^2}\approx 153\gev\;.
\end{equation}
Only off-shell effects and further radiation give access to the phase space region above $153\gev$. For the presented calculation we see that already
the LO prediction exceeds this boundary and NLO corrections are large in this region. The off-shell contributions to the total cross section are small.
However, they are important in the determination of the endpoint of the distribution because they smear this sharp boundary. The knowledge of the endpoint
allows to infer the top-quark mass~\cite{Chatrchyan:2013boa,Aad:2015nba}.
%
\section{Conclusions}
In this proceeding we reported the first calculation of NLO QCD corrections for\\ \proc{}, where $W, Z$ and top quark off-shell effects have been
taken into account. The NLO QCD corrections turn out to be moderate for the total cross section as well as for differential distributions.
For instance, the total cross section is reduced by $-13\%$, while the remaining scale uncertainty is at the level of $20\%$.

The study of this process will be extended in the future by a detailed comparison of fixed and dynamic renormalization and factorization scales.
Also a comparison between the four and the five-flavour scheme is desirable, since the uncertainties connected to the chosen scheme can reach up
to $10\%$, as reported in Ref.~\cite{Bevilacqua:2013taa}. Finally, the calculation can be used to improve predictions for the top-quark mass parameter
extraction using an alternative method~\cite{Alioli:2013mxa,Aad:2015waa}.

\textbf{Acknowledgement:} This work has been supported by the DFG under the Grant No. WO 1900/1-1 --
\textit{Signals and Backgrounds Beyond Leading Order. Phenomenological studies for the LHC}.


\end{document}